\documentclass[11pt]{article}
\usepackage{amsfonts,slashed}
\usepackage{arydshln}
\usepackage[greek,english]{babel}
\usepackage{cancel}
\usepackage{upgreek}
\usepackage{float}
\usepackage{url}
\usepackage{latexsym}
\usepackage{amsfonts}
\usepackage{amsmath}
\usepackage{epsfig}
\usepackage{latexsym,amssymb}
\usepackage{amsmath,amssymb,amsthm}
\usepackage{fontenc}
\usepackage{graphicx,tikz}
\usepackage{multirow}
\usepackage{xfrac}
\usepackage[T1]{fontenc}
\usepackage{textcomp}
\usepackage{lmodern}
\usepackage{stmaryrd}
\usepackage{cite}

\usepackage{stmaryrd}

\usepackage[margin=20pt,small]{caption}
\usepackage{subcaption}

\usepackage{ifpdf}
\ifx\pdfoutput\undefined
   \pdffalse
   \usepackage{cite}
 \else
   \pdfoutput=1
   \pdftrue
\fi
\DeclareFontFamily{OMS}{rsfs}{\skewchar\font'60}
\DeclareFontShape{OMS}{rsfs}{m}{n}{<-5>rsfs5 <5-7>rsfs7 <7->rsfs10 }{}
\DeclareSymbolFont{rsfs}{OMS}{rsfs}{m}{n}
\DeclareSymbolFontAlphabet{\Scr}{rsfs}

\numberwithin{equation}{section}
\def\be{\begin{equation}}
\def\ee{\end{equation}}
\def\ba{\begin{array}}
\def\ea{\end{array}}

\newcommand{\bea}{\begin{eqnarray}}
\newcommand{\eea}{\end{eqnarray}}

\newcommand{\ft}[2]{{\textstyle\frac{#1}{#2}}}

\def\={~=~}
\def\*{{}^*}

\usepackage[shadow,textwidth=2.7cm]{todonotes}
\usepackage{ifthen}
\reversemarginpar
\newcounter{todocounter}

\colorlet{fccolor}{blue!40!white}

\newcommand{\fcinline}[2][]{
  \ifthenelse { \equal {#1} {} }
    { \def\temp {#2} }  
    { \def\temp {#1} }   
  \refstepcounter{todocounter}\todo[color=fccolor,inline,caption={\textbf{\thetodocounter. FC} \temp}]{\textbf{\thetodocounter. FC:} #2}{}}

\colorlet{hscolor}{orange!20!white}

\newcommand{\hsinline}[2][]{
  \ifthenelse { \equal {#1} {} }
    { \def\temp {#2} }  
    { \def\temp {#1} }   
  \refstepcounter{todocounter}\todo[color=hscolor,inline,caption={\textbf{\thetodocounter. HS} \temp}]{\textbf{\thetodocounter. HS:} #2}{}}

\def\={~=~}
\def\*{{}^*}

\newcommand{\hodge}{\star}

\begin{document}
\begin{titlepage}
\vfill
\begin{flushright}
\end{flushright}

\bigskip

\begin{center}
	{\LARGE \bf Consistent sphere reductions of gravity \\[4mm]
	to two dimensions
	}\\[1cm]
	
	{\large\bf Franz Ciceri\,$^{a}{\!}$
		\footnote{\tt franz.ciceri@aei.mpg.de}, Henning Samtleben\,${}^{b,c}{\!}$
		\footnote{\tt henning.samtleben@ens-lyon.fr} \vskip .8cm}	
	{\it  $^{a}$ Max Planck Institute for Gravitational Physics (Albert Einstein Institute),\\ Am M\"uhlenberg 1, 14476 Potsdam, Germany}\\[2ex] \ 
		{\it ${}^b$ ENSL, CNRS, Laboratoire de physique, F-69342 Lyon, France}\\[2ex] \ 
	{\it  $^{c}$ Institut Universitaire de France (IUF)}\\ \ \\
	
\end{center}
\vfill

\begin{center}
	\textbf{Abstract}
	
\end{center}
\begin{quote}

Consistent reductions of higher-dimensional (matter-coupled)
gravity theories on spheres have been constructed and classified 
in an important paper by Cveti\v{c}, L\"u and Pope.
We close a gap in the classification and study the case when the resulting lower-dimensional
theory is two-dimensional. 
We construct the consistent reduction of Einstein-Maxwell-dilaton gravity on a $d$-sphere $S^d$
to two-dimensional dilaton-gravity
coupled to a gauged sigma model with target space ${\rm SL}(d+1)/{\rm SO}(d+1)$.
The truncation contains solutions of type AdS$_2\times \Sigma_d$ where the internal space $\Sigma_d$ is a deformed sphere.
In particular, the construction includes the consistent truncation around the near-horizon geometry of the boosted Kerr string.
In turn, we find that an AdS$_2\times S^d$ background with the round $S^d$ within a consistent truncation
requires $d>3$ and an additional cosmological term in the higher-dimensional theory.

\end{quote}
\vfill
\setcounter{footnote}{0}

\end{titlepage}

\tableofcontents \noindent {}


\section{Introduction} 
\label{sec:Intro}

Consistent truncations of higher-dimensional gravity theories have a long history
nourished by the emergence of extra dimensions in supergravity and string theory.
Explicitly, this is the question if a gravitational theory can be truncated to a finite set
of fields whose dynamics is described by a lower-dimensional action, such that any solution of
the lower-dimensional theory can be uplifted to a solution of the original, higher-dimensional theory.
In terms of the infinite towers of Kaluza-Klein fluctuations around a given background, this corresponds to
the truncation to a finite number of such fluctuations which is consistent at the full non-linear level.
In general, non-linear products of the fields that are being retained will act as sources for the truncated fields,
thereby rendering the truncation inconsistent. Consistent truncations are not low energy effective 
field theories since retained fields may have masses comparable to the truncated fields. Yet, they provide very
powerful tools for the construction of higher-dimensional exact solutions as well as for the applicability of supergravity 
techniques for holographic dualities.

Unlike for toroidal reductions which retain precisely the singlets under the ${\rm U}(1)^d$ isometry of the torus $T^d$
such that consistency follows from a simple group-theoretic argument, the question becomes much more involved
for non-trivial internal spaces.
In an important paper, Cveti\v{c}, L\"u and Pope have classified and explicitly constructed the consistent reductions
of (matter-coupled) gravity theories on spheres $S^d$ \cite{Cvetic:2000dm},
that retain all the Yang-Mills fields of ${\rm SO}(d+1)$, gauging the full isometry group of the sphere.
A strong necessary condition for the existence of such a consistent truncation can be found from the toroidal reduction
of the relevant higher-dimensional theory. Its global symmetry group ${\rm G}$ must accommodate an ${\rm SO}(d+1)$ subgroup
such that gauging of the latter describes the theory obtained from reduction on the sphere. In general, this requires some
symmetry enhancement that only occurs for particular matter content and couplings of the higher-dimensional theory.
For example, a straightforward counting argument along these lines \cite{Cvetic:2000dm} shows that the reduction of $D$-dimensional 
gravity coupled to a single $d$-form field strength $F_{(d)}$ on the sphere $S^d$ can only be consistent for
\begin{equation}
(D,d)\in\left\{(11,7), \,(11,4), \,(10,5) \right\}
\;.
\label{eq:Dpex}
\end{equation}
These are precisely the reductions, realized for $D=11$ supergravity on $S^7$ \cite{deWit:1986iy,deWit:2013ija,Nicolai:2011cy,Godazgar:2013pfa},
$D=11$ supergravity on $S^4$ \cite{Nastase:1999kf}, and IIB supergravity on $S^5$ \cite{Baguet:2015sma}, respectively. 
They describe consistent truncations around AdS$_{D-d}\times S^d$ backgrounds to the maximal supergravity multiplet,
and as such have played important roles in the AdS/CFT dualities.

Further possibilities for such consistent sphere reductions exist when the higher-dimensional theory
carries an additional dilaton field, i.e.\ is described by a $D$-dimensional Lagrangian 
\be
{\cal L}_D = \hat R\; {\hat \star1} - \tfrac12\, {\hat \star d\hat\phi}\wedge
d\hat\phi - \tfrac12 e^{-a\, \hat\phi}\, {\hat \star\hat F_{{(d)}}}\wedge
\hat F_{{(d)}}\,,
\label{eq:LDintro}
\ee
with a particular value for the constant $a=a(D,d)$\,. As shown in \cite{Cvetic:2000dm}, the symmetry enhancement
required for existence of a consistent reduction on $S^d$ arises for the four families of theories with
\begin{equation}
(D,d)\in\big\{ (D,2), (D,3), (D,D-3), (D,D-2) \big\}
\;,
\label{eq:Dp23}
\end{equation}
where the latter two are the Hodge duals of the first two.
Furthermore, in \cite{Cvetic:2000dm} the consistent reductions on $S^2$, $S^3$, $S^{D-3}$, 
corresponding to the first three families of (\ref{eq:Dp23}) were explicitly constructed.
The last case of (\ref{eq:Dp23}), i.e.\ reduction of the theory (\ref{eq:LDintro}) on a sphere $S^{D-2}$ has been left aside,
mostly because the resulting theory is a two-dimensional gravity theory in which many of the generic structures degenerate.
For instance, the theory enjoys an additional global Weyl symmetry and cannot be cast into the canonical Einstein frame.
Furthermore, reductions on $T^{(D-2)}$ lead to ungauged gravity theories in two dimensions, which come with an infinite-dimensional symmetry enhancement,
generalising the affine Geroch group, originally discovered in the dimensional reduction of $D=4$ general relativity \cite{Geroch:1972yt}. 

In this paper, we complete the classification of \cite{Cvetic:2000dm} and explicitly construct the
consistent truncation of the theory (\ref{eq:LDintro}) on a sphere $S^{D-2}$\,. 
The resulting theories are two-dimensional dilaton gravity theories which may carry AdS$_2$ vacua that uplift to higher-dimensional
AdS$_2\times S^d$ backgrounds. As such they may provide important tools to describe fluctuations around the near-horizon geometry of (near-)extremal black holes. This is especially relevant for the study of such black holes in the context of AdS$_2$ holography, see for instance~\cite{Cvetic:2016eiv,Sarosi:2017ykf,Nayak:2018qej,Castro:2018ffi,Castro:2021wzn,Aniceto:2021xhb}.

We present the full non-linear reduction ansatz for the fields of (\ref{eq:LDintro}) on a sphere $S^{D-2}$. 
The matter sector of the resulting two-dimensional theories is a gauged sigma model with target space ${\rm SL}(d+1)/{\rm SO}(d+1)$
and a scalar potential. The gauge fields appear with a two-dimensional YM term. Consistency of the truncation requires that the constant $a$ in
(\ref{eq:LDintro}) must take a particular value. As it turns out, for this value the $D$-dimensional theory itself can be obtained by circle reduction of pure gravity in $(D+1)$ dimensions.
Accordingly, we also work out the uplift of the two-dimensional theory to $(D+1)$ dimensions. For $D=10$, our result coincides with the pure gravity sector of the ansatz constructed in \cite{Bossard:2022wvi,Bossard:2023jid}, that describes the consistent truncation of $D=11$ supergravity on $S^8\times S^1$ using affine exceptional field theory \cite{Bossard:2017aae,Bossard:2018utw,Bossard:2021jix}.

We construct a number of solutions of the two-dimensional theories, mostly restricting to solutions with constant scalars and dilaton.
We find multi-parameter families of such solutions living in the truncation of the two-dimensional theory to singlets under the ${\rm U}(1)^{[\frac{d+1}{2}]}$ Cartan subgroup of ${\rm SO}(d+1)$. They naturally generalize the solutions found in \cite{Anabalon:2013zka} for the case of $S^8$. Interestingly, we find that all such solutions necessarily break ${\rm SO}(d+1)$, implying that the corresponding higher-dimensional AdS$_2\times S^d$ backgrounds all involve deformations of the round $S^d$ sphere.
Notably, these solutions include the $D=5$ near horizon geometry of the boosted Kerr string \cite{Frolov:1987rj,Kunduri:2007vf}.

Embedding of an AdS$_2\times S^d$ background with the round $S^d$ on the other hand requires the addition of
a cosmological term 
\begin{equation}
{\cal L}_{D,m} = -\frac12\,m^2\,e^{b\hat\phi}\,{\hat\hodge1}\;,
\label{eq:ccIntro}
\end{equation}
to the $D$-dimensional theory (\ref{eq:LDintro}), with the constant $b$ tuned to a particular value. Still, this turns out to be possible only for $D>5$.
In contrast, the $D=4$ and $D=5$ theories admit dS$_2\times S^2$ and Mink$_2\times S^3$ backgrounds, respectively,
within their consistent truncations.

The rest of this paper is organized as follows. In section~\ref{sec:S2S3}. we start by reviewing the results of \cite{Cvetic:2000dm} on consistent 
$S^2$ and $S^3$ reductions from $D$ dimensions. We describe how to properly extrapolate the constructions to $D=4$ and $D=5$, respectively,
such that the resulting theories are two-dimensional. In section~\ref{sec:consistentS} we generalize the structure to arbitrary dimensions and 
construct the consistent truncation of the theory (\ref{eq:LDintro}) on a sphere $S^{D-2}$\,. 
We describe the further uplift to pure gravity in $(D+1)$ dimensions and the inclusion of a cosmological term (\ref{eq:ccIntro}).
In section~\ref{sec:solutions}, we study solutions of the two-dimensional theories and their uplift to $D$ dimensions.
We close with some conclusions in section~\ref{sec:conclusions} where we also discuss the symmetries underlying the presented constructions.

\section{Consistent $S^2$ and $S^3$ reductions} 
\label{sec:S2S3}

\subsection{Review of previous results}

We start by reviewing the results of \cite{Cvetic:2000dm} on consistent 
$S^d$ reductions from $D$ dimensions for $d=2, 3$. The $D$-dimensional theories
are of the type (\ref{eq:LDintro}), i.e.\ an Einstein-Maxwell dilaton system for $d=2$,
and the bosonic string with Kalb-Ramond field and a dilaton for $d=3$.

Let us first describe the $S^2$ case: starting from the 
Einstein-Maxwell dilaton system\footnote{
Following \cite{Cvetic:2000dm}, we use Hodge star conventions 
$\hat{\star}\alpha \wedge \beta = \langle \alpha,\beta\rangle \,\hat\omega_D= \langle \alpha,\beta\rangle \,{\hat\hodge1}$.}
\begin{equation}
{\cal L}_D = \hat R\; {\hat \hodge1} - 
\frac12 \,{\hat \hodge d\hat\phi}\wedge d\hat\phi  
-\frac12 \, e^{-\sqrt{\frac{2\,(D-1)}{D-2}}\, \hat\phi}\, {\hat \hodge\hat F_{(2)}}\wedge \hat F_{(2)}
\;,
\label{eq:S2L}
\end{equation}
in $D$ dimensions, the consistent reduction ansatz on an internal $S^2$ is given by
\bea
d\hat s_D^2 &=& Y^{\frac1{D-2}}\, \Big(\Delta^{\frac1{D-2}}\, ds_{D-2}^2 
+ g^{-2}\, \Delta^{-\frac{D-3}{D-2}}\, T^{-1}_{ij}\, 
{\cal D}\mu^i\, {\cal D}\mu^j\Big)
\;,\nonumber\\
\hat F_{(2)} &=& \tfrac12 \epsilon_{ijk}\, \Big( g^{-1}\, U\, \Delta^{-2}\, 
\mu^i\, {\cal D}\mu^j\wedge {\cal D}\mu^k - 2 g^{-1}\, \Delta^{-2}\,
{\cal D}\mu^i\wedge {\cal D} T_{j\ell}\, T_{k m}\, \mu^\ell\, \mu^m
\nonumber\\
&&\qquad\qquad  - \Delta^{-1}\, F_{(2)}^{ij}\, T_{k\ell}\, \mu^\ell \Big)
\;,\nonumber\\
e^{\sqrt{\frac{2(D-2)}{D-1}}\, \hat\phi} &=& \Delta^{-1}\, 
Y^{\frac{D-3}{D-1}}
\;.
\label{eq:S2Dansatz}
\eea
Here, the $\mu^i$ are the embedding coordinates of the $S^2$ sphere
\begin{equation}
\mu^i\mu^i=1\;,\quad i=1, 2, 3\;,
\end{equation}
the symmetric and positive definite matrix $T_{ij}$ carries the lower-dimensional scalar fields, and
\begin{equation}
\Delta\equiv{T}_{ij}\,\mu^i\mu^j\;,\quad
Y\equiv{\rm det}\,T_{ij}\;,\quad
U\equiv2\,{T}_{ik}{T}_{jk}\,\mu^i\mu^j-\Delta{T}_{ii}
\;.
\label{eq:DTU}
\end{equation}
The inverse sphere radius $g$ appears as a coupling constant for the ${\rm SO}(3)$ covariant derivatives
\begin{equation}
{\cal D} \mu^i = d\mu^i + g\,A_{(1)}^{ij}\,\mu^j\;,\qquad
{\cal D}  T_{ij} =
d  T_{ij} + g\,A_{(1)}^{ik}  T_{kj} + g\,A_{(1)}^{jk}  T_{ik} 
\;,
\label{eq:covD}
\end{equation}
with the associated Yang-Mills field strength given by
\begin{equation}
F_{(2)}^{ij}=dA_{(1)}^{ij} + g\,A_{(1)}^{ik} \wedge A_{(1)}^{kj} 
\;.
\label{eq:F2YM}
\end{equation}
It has been shown in \cite{Cvetic:2000dm} that plugging the reduction ansatz (\ref{eq:S2Dansatz}) into the 
field equations obtained from (\ref{eq:S2L}), all the dependence on the sphere coordinates consistently factors out
and the equations reduce to $(D-2)$-dimensional field equations which are obtained from variation of the 
$(D-2)$-dimensional Lagrangian
\bea
{\cal L}_{D-2} &=& R\, {\hodge1} - \tfrac{D-4}{3(D-1)}\, Y^{-2}\,
{\hodge dY}\wedge dY - \tfrac14 \tilde T^{-1}_{ij}\, {\hodge{\cal D} \tilde T_{jk}}\wedge
\tilde T^{-1}_{k\ell}\, {\cal D}\tilde T_{\ell i} \nonumber\\
&& -\tfrac1{4}\, Y^{-2/3}\,  \tilde T^{-1}_{ik}\,
\tilde T^{-1}_{j\ell}\, {\hodge F_{(2)}^{ij}}\wedge F_{(2)}^{k\ell}
- \tfrac12 g^2\, Y^{2/3}\, \Big(2\, \tilde T_{ij}\, \tilde T_{ij} 
  - (\tilde T_{ii})^2 \Big)\, {\hodge1}\,,
  \label{eq:L2DredS2}
\eea
where the matrix $T_{ij}$ is parametrized as $T_{ij}= Y^{1/3}\, \tilde T_{ij}$ such that ${\rm det}\,\tilde{T}_{ij}=1$.  
The computation of the $D$-dimensional field equations also exploits the explicit form of the Hodge dual of $\hat{F}_{(2)}$ from (\ref{eq:S2Dansatz})
which is found to be given by
\bea
e^{-\sqrt{\frac{2(D-1)}{D-2}}\, \hat\phi}\, \hat \hodge  \hat F_{(2)} &=& 
-g\, U\, \omega_{D-2} + g^{-1}\, T^{-1}_{ij}\, 
{\hodge {\cal D} T_{jk}}\wedge (\mu^k\, {\cal D} \mu^i) 
\nonumber\\
&&
-\tfrac1{2}\,g^{-2}  T^{-1}_{ik}\, T^{-1}_{j\ell}\, 
{\hodge  F_{(2)}^{ij}}\wedge {\cal D} \mu^k\wedge
{\cal D}\mu^\ell
\;.
\label{eq:F2Hodge}
\eea
Throughout, $ \hat \hodge $ refers to the Hodge star in $D$ dimensions, while $\hodge $ refers to the Hodge star in $(D-2)$ dimensions.
The volume form of the $(D-2)$ dimensional metric is denoted by $\omega_{D-2}$.


Consistent truncations on $S^3$ are constructed in an analogous way.
Starting from the Lagrangian of the bosonic string
\begin{equation}
{\cal L}_D = \hat R\; {\hat \hodge 1} - 
\tfrac12 {\hat \hodge d\hat\phi}\wedge d\hat\phi  
-\tfrac12 e^{-\sqrt{\frac8{D-2}}\, \hat\phi}\, {\hat \hodge \hat F_{(3)}}\wedge \hat F_{(3)}
\;,
\label{eq:S3L}
\end{equation}
in $D$ dimensions, the consistent reduction ansatz on an internal $S^3$ is given by \cite{Cvetic:2000dm}
\bea
d\hat s_D^2 &=& Y^{\frac1{D-2}}\, \Big( \Delta^{\frac2{D-2}}\, ds_{D-3}^2 
+ g^{-2}\, \Delta^{-\frac{D-4}{D-2}}\,  T^{-1}_{ij}\, 
{\cal D}\mu^i\, {\cal D}\mu^j\Big) \,,\label{metans}\nonumber\\
\hat F_{(3)} &=& F_{(3)} + \ft16\, \epsilon_{i_1 i_2 i_3 i_4}\, 
\Big( g^{-2}\, U\, \Delta^{-2}\, 
  {\cal D}\mu^{i_1}\wedge {\cal D}\mu^{i_2} \wedge {\cal D}\mu^{i_3}\, 
\mu^{i_4} \nonumber\\
&&- 3g^{-2}\, \Delta^{-2} \, 
D\mu^{i_1} \wedge {\cal D}\mu^{i_2}\wedge {\cal D} T_{i_3 j}\,
T_{i_4 k}\, \mu^j\, \mu^k  - 3g^{-1}\, \Delta^{-1}\, F_{(2)}^{i_1 i_2} \wedge
{\cal D}\mu^{i_3}\, T_{i_4 j}\, \mu^j \Big)\,,
\nonumber\\
e^{\sqrt{\frac{D-2}{2}}\, \hat\phi} &=& \Delta^{-1}\, Y^{\frac{D-4}{4}}\,.
\label{eq:S3Dansatz}
\eea
Here, the $\mu^i$ are the embedding coordinates of the $S^3$ sphere
\begin{equation}
\mu^i\mu^i=1\;,\quad i=1, \dots, 4\;,
\end{equation}
while the quantities $\Delta$, $Y$, and $U$ and the covariant derivatives are defined as in (\ref{eq:DTU}) and (\ref{eq:covD}) above.
On top of the ${\rm SO}(4)$ field strength (\ref{eq:F2YM}), the lower-dimensional theory carries a two-form gauge potential $B_{(2)}$
whose field strength is defined as
\begin{equation}
F_{(3)} = dB_{(2)} + \tfrac18 \epsilon_{ijk\ell}\, A_{(1)}^{ij}\wedge F_{(2)}^{k\ell}\,.
\label{eq:F3}
\end{equation}
It has been shown in \cite{Cvetic:2000dm} that plugging the reduction ansatz (\ref{eq:S3Dansatz}) into the 
field equations obtained from (\ref{eq:S3L}), all the dependence on the sphere coordinates consistently factors out
and the equations reduce to $(D-3)$-dimensional field equations which are obtained from variation of the 
$(D-3)$-dimensional Lagrangian
\bea
{\cal L}_{D-3} &=& R\, {\hodge 1} - \tfrac{D-5}{16}\, Y^{-2}\,{\hodge dY}\wedge dY - \tfrac14 \,\tilde T^{-1}_{ij}\, {\hodge {\cal D} \tilde T_{jk}}\wedge
\tilde T^{-1}_{k\ell}\, {\cal D}\tilde T_{\ell i} 
 -\tfrac12\,Y^{-1}\,\hodge F_{(3)}\wedge F_{(3)}
 \nonumber\\
&&
-\tfrac1{4}\, Y^{-1/2}\,  \tilde T^{-1}_{ik}\,
\tilde T^{-1}_{j\ell}\, {\hodge  F_{(2)}^{ij}}\wedge F_{(2)}^{k\ell}
- \tfrac12 \,g^2\, Y^{1/2}\, \Big(2\, \tilde T_{ij}\, \tilde T_{ij} 
  - (\tilde T_{ii})^2 \Big)\, {\hodge 1}\,,
  \label{eq:L2DredS3}
\eea
where the matrix $T_{ij}$ is parametrized as $T_{ij}= Y^{1/4}\, \tilde T_{ij}$ such that ${\rm det}\,\tilde{T}_{ij}=1$.  
The computation of the $D$-dimensional field equations also exploits the explicit form of the Hodge dual of $\hat{F}_{(3)}$ from (\ref{eq:S3Dansatz})
which is found to be given by
\bea
e^{-\sqrt{\frac{8}{D-2}}\, \hat\phi}\, {\hat \hodge  \hat F_{(3)}} &=& \tfrac1{6} \,g^{-3}\,
\epsilon_{ijk\ell}\,  Y^{-1} {\hodge  F_{(3)}}\wedge \mu^i\, {\cal D}\mu^j\wedge {\cal D}\mu^k
\wedge {\cal D} \mu^\ell -g\, U\, \omega_{D-3}
\label{eq:F3Hodge}
\\
&& + g^{-1}\, T^{-1}_{ij}\, 
{\hodge {\cal D} T_{jk}}\wedge (\mu^k\, {\cal D}\mu^i) 
-\tfrac1{2}  g^{-2}\,  T^{-1}_{ik}\, T^{-1}_{j\ell}\, 
{\hodge  F_{(2)}^{ij}}\wedge {\cal D}\mu^k\wedge
{\cal D}\mu^\ell\,.
\nonumber
\eea

\subsection{Reductions to two dimensions}

We have in the previous section reviewed the results of \cite{Cvetic:2000dm} on the
consistent truncations of $D$-dimensional theories (\ref{eq:LDintro}) on $S^2$ and $S^3$.
Although derived for higher dimensions, in principle, the entire construction goes through 
even in the case when the resulting theory
is two-dimensional, i.e.\ for $D=4$ on $S^2$ and $D=5$ on $S^3$.
In particular, none of the reduction ansaetze (\ref{eq:S2Dansatz}), (\ref{eq:S3Dansatz}),
diverges at these values for $D$.
However, the resulting Lagrangians (\ref{eq:L2DredS2}), (\ref{eq:L2DredS3}), when evaluated in 
two dimensions, give rise to equations of motion that kill all of their dynamical content
as a consequence of the two-dimensional Einstein equations. Consequently, the two-dimensional case
was left apart in \cite{Cvetic:2000dm}. Another indication that the Lagrangians (\ref{eq:L2DredS2}), (\ref{eq:L2DredS3}), do not
get along well with a two-dimensional space-time is the fact that the reduction of the higher-dimensional Einstein-Hilbert term
in general yields the two-dimensional Einstein-Hilbert term only up to a dilatonic prefactor. While in generic dimensions
this factor can be removed by a Weyl transformation of the metric, this is no longer the case in two dimensions. A generic reduction
to two dimensions will thus produce dilaton gravity rather than pure two-dimensional gravity.

In short, while the results of the previous section are still valid for the truncation to a two-dimensional theory, the resulting theory
does not describe a dynamical subsector of the higher-dimensional theory.
What we show here is how the situation can be remedied by properly redefining the fields in the above structures, before
extrapolating the construction to two dimensions.

Consider first the $S^2$ reduction of the theory (\ref{eq:S2L}). 
Starting in general dimension $D$, we may redefine the $(D-2)$-dimensional fields appearing 
in the ansatz (\ref{eq:S2Dansatz}) as
\begin{equation}
Y \equiv \rho^{-\frac{\sqrt{3(D-1)(D-3)}}{D-4}}
\;,\quad
g_{\mu\nu} = \rho^{\frac{2}{D-4}}\,\tilde{g}_{\mu\nu}
\;.
\label{eq:rescalingS2}
\end{equation}
Even though this rescaling is clearly singular at $D=4$, the resulting ansatz (\ref{eq:S2Dansatz}) still has a smooth limit $D\rightarrow4$.
Namely, defining
\begin{equation}
{T}_{ij} = Y^{1/3}\,\tilde{T}_{ij}\;,\quad
\Delta =  Y^{1/3}\,\tilde{\Delta}\;,\quad
U=Y^{2/3}\,\tilde{U}\;,
\end{equation}
in order to extract the $Y$-dependence of the various quantities, the ansatz (\ref{eq:S2Dansatz}) takes the form
\bea
d\hat s_D^2 &=&   \rho^{\frac{6(D-2)-4\sqrt{3(D-1)(D-3)}}{3(D-2)(D-4)}}  \tilde\Delta^{\frac1{D-2}}\, d\tilde{s}_{D-2}^2 
+ g^{-2}\, \rho^{\frac{2\sqrt{3(D-1)(D-3)}}{3(D-2)}}  \tilde\Delta^{-\frac{D-3}{D-2}}\, \tilde{T}^{-1}_{ij}\, 
{\cal D}\mu^i\, {\cal D}\mu^j
\;,\nonumber\\
e^{\sqrt{\frac{2(D-2)}{D-1}}\, \hat\phi} &=&  
\rho^{\frac{-2\sqrt{D-3}}{\sqrt{3(D-1)}}}\,\tilde\Delta^{-1}
\;,\nonumber\\[1ex]
\hat F_{(2)} &=& \frac1{2g\,\tilde\Delta^{2}} \epsilon_{ijk}\, \Big(\tilde U\, 
\mu^i\, {\cal D}\mu^j\wedge {\cal D}\mu^k - 2 \,
{\cal D}\mu^i\wedge {\cal D} \tilde T_{j\ell}\, \tilde T_{k m}\, \mu^\ell\, \mu^m
- g\,\tilde\Delta\, F_{(2)}^{ij}\, \tilde T_{k\ell}\, \mu^\ell \Big)
\;,
\label{eq:truncD4D}
\eea
where now $d\tilde{s}_{D-2}^2$ refers to the new metric $\tilde{g}_{\mu\nu}$ from (\ref{eq:rescalingS2}).
So far, this is merely a rewriting of the ansatz (\ref{eq:S2Dansatz}) in different variables, but upon taking the limit $D\rightarrow4$,
we obtain the smooth limit of $D$-dimensional metric and dilaton to
\bea
d\hat s_4^2 &=&   \rho^{-1/3}  \tilde\Delta^{1/2}\, ds_{2}^2 
+ g^{-2}\, \rho\, \tilde\Delta^{-1/2}\, \tilde{T}^{-1}_{ij}\, 
{\cal D}\mu^i\, {\cal D}\mu^j
\;,\nonumber\\
e^{\sqrt{3}\, \hat\phi} &=&  
\rho^{-1}\,\tilde\Delta^{-3/2}
\;,
\label{eq:truncD4}
\eea
which differs from the naive $D\rightarrow4$ limit of (\ref{eq:S2Dansatz}) by  
the non-trivial powers of $\rho$.
Furthermore, the expression for $\hat F_{(2)}$ in (\ref{eq:truncD4D}) has no explicit $Y$ or ${g}_{\mu\nu}$ dependence
and retains its form for $D\rightarrow4$.
Accordingly, the resulting Lagrangian (\ref{eq:L2DredS2}) after rescaling (\ref{eq:rescalingS2})
also yields a non-trivial smooth limit to $D\rightarrow4$, given by
\begin{equation}
{\cal L}_{2} =
 \rho\,\tilde{R}\,{\tilde\hodge 1}
+\tfrac14\,\rho\,{\tilde\hodge {\cal D}} \tilde{T}_{ij}^{-1}\wedge {\cal D} \tilde{T}_{ij} 
-\tfrac14\,\rho^{\frac73}\,\tilde{T}_{ij}^{-1}\tilde{T}_{k\ell}^{-1} \,{\tilde\hodge F_{(2)}^{ik}}\wedge F_{(2)}^{j\ell} 
- \tfrac12\,g^2\,\rho^{-\frac13}\left(2\,\tilde{T}_{ij}\tilde{T}_{ij}-\tilde{T}_{ii}^2 \right) {\tilde\hodge 1}
\;,
\label{eq:L2D2S2}
\end{equation}
again differing from the naive $D\rightarrow4$ limit of (\ref{eq:L2DredS2}) by the non-trivial powers of $\rho$.
Again, $\tilde{R}$ and $\tilde\hodge $ refer to the new metric $\tilde{g}_{\mu\nu}$ from (\ref{eq:rescalingS2}).
We may now forget about the singular rescaling (\ref{eq:rescalingS2}), and directly prove that the regular reduction ansatz 
(\ref{eq:truncD4}) defines a consistent truncation of the 4-dimensional Einstein-Maxell dilaton system (\ref{eq:S2L}) on $S^2$.
We give the details in section~\ref{sec:reduction} below, where we discuss the general $S^{D-2}$ truncation.
The resulting two-dimensional theory is given by the Lagrangian (\ref{eq:L2D2S2}).
For later convenience, we also note that the dual field strength (\ref{eq:F2Hodge}) for $D=4$ 
and in terms of the fields (\ref{eq:rescalingS2}) is found to be
\begin{align}
\hat{\cal F}_{(2)} \equiv&\;\; e^{-\sqrt{3}\, \hat\phi}\, \hat \hodge  \hat F_{(2)} 
\label{eq:F2HodgeD4}\\
=&\;
-g\,\rho^{-1/3}\,\tilde{U}\, \tilde\omega_{(2)} + g^{-1}\rho\, \tilde{T}^{-1}_{ij}\, 
{\tilde\hodge {\cal D} \tilde{T}_{jk}}\wedge (\mu^k\, {\cal D} \mu^i) 
-\tfrac12\,g^{-2}\, \rho^{7/3}\, \tilde{T}^{-1}_{ik}\, \tilde{T}^{-1}_{j\ell}\, 
{\tilde\hodge  F_{(2)}^{ij}} \, {\cal D} \mu^k\wedge {\cal D}\mu^\ell
\nonumber
\;.
\end{align}
In terms of this dual field strength, the original $D=4$ theory (\ref{eq:S2L}) can be equivalently rewritten as
\bea
{\cal L}_4 &=& \hat R\; {\hat \hodge 1} - 
\tfrac12 \,{\hat \hodge d\hat\phi}\wedge d\hat\phi  
-\tfrac12 \,e^{\sqrt{3}\, \hat\phi}\, {\hat \hodge \hat {{\cal F}}_{(2)}}\wedge \hat {{\cal F}}_{(2)}
\;.
\label{eq:L4dual}
\eea


Similarly, we can extend the general $S^3$ reduction of the previous section to the particular case $D=5$ in which the
resulting theory becomes two-dimensional. Starting from the $S^3$ reduction of the theory (\ref{eq:S3L}) in general dimension $D$, we may redefine the $(D-3)$-dimensional fields appearing 
in the ansatz (\ref{eq:S2Dansatz}) as
\begin{equation}
Y \equiv \rho^{-\frac{4\sqrt{D-4}}{D-5}}
\;,\quad
g_{\mu\nu} =  \rho^{\frac{2}{D-5}}\,\tilde{g}_{\mu\nu}
\;.
\label{eq:rescalingS3}
\end{equation}
This rescaling is singular at $D=5$, but the resulting ansatz (\ref{eq:S3Dansatz}) still has a smooth limit $D\rightarrow5$.
Defining
\begin{equation}
{T}_{ij} = Y^{1/4}\,\tilde{T}_{ij}\;,\quad
\Delta =  Y^{1/4}\,\tilde{\Delta}\;,\quad
U=Y^{1/2}\,\tilde{U}\;,
\end{equation}
in order to extract the $Y$-dependence of the various quantities, the ansatz (\ref{eq:S3Dansatz}) takes the form
\bea
d\hat s_D^2 &=& \rho^{\frac{D-2-3\sqrt{D-4}}{(D-2)(D-5)}}\,\tilde\Delta^{\frac2{D-2}}\,d\tilde{s}_{D-3}^2 
+ g^{-2}\, \rho^{\frac{2\sqrt{D-4}}{D-2}} \tilde\Delta^{-\frac{D-4}{D-2}}\,  \tilde{T}^{-1}_{ij}\, 
{\cal D}\mu^i\, {\cal D}\mu^j \label{eq:truncD5D} \\
e^{\sqrt{(D-2)/2}\, \hat\phi} &=& \rho^{-\sqrt{D-4}} \, \tilde\Delta^{-1}\, \,,
\nonumber\\
\hat F_{(3)} &=& F_{(3)} + \ft16\, \epsilon_{i_1 i_2 i_3 i_4}\, 
\Big( g^{-2}\,  \tilde{U}\, \tilde{\Delta}^{-2}\, 
  {\cal D}\mu^{i_1}\wedge {\cal D}\mu^{i_2} \wedge {\cal D}\mu^{i_3}\, 
\mu^{i_4} \label{eq:FSd=3}\\
&&- 3g^{-2}\, \tilde{\Delta}^{-2} \, 
D\mu^{i_1} \wedge {\cal D}\mu^{i_2}\wedge {\cal D} \tilde{T}_{i_3 j}\,
\tilde{T}_{i_4 k}\, \mu^j\, \mu^k  - 3g^{-1}\, \tilde{\Delta}^{-1}\, F_{(2)}^{i_1 i_2} \wedge
{\cal D}\mu^{i_3}\, \tilde{T}_{i_4 j}\, \mu^j \Big)
\;,\nonumber
\eea
where now $d\tilde{s}_{D-3}^2$ refers to the new metric $\tilde{g}_{\mu\nu}$ from (\ref{eq:rescalingS3}).
The limit $D\rightarrow5$ is smooth and yields 
\bea
d\hat s_5^2 &=& \rho^{-1/3}\,\tilde\Delta^{2/3}\,ds_{2}^2 
+ g^{-2}\, \rho^{2/3} \tilde\Delta^{-1/3}\,  \tilde{T}^{-1}_{ij}\, 
{\cal D}\mu^i\, {\cal D}\mu^j \;,
\nonumber\\
e^{\sqrt{8/3}\, \hat\phi} &=& \rho^{-4/3} \, \tilde\Delta^{-4/3}
\;,
\label{eq:truncD5}
\eea
for the $D=5$ metric and dilaton, which differs from the naive $D\rightarrow5$ limit of (\ref{eq:S3Dansatz}) by  
the non-trivial powers of $\rho$.
The expression for $\hat F_{(3)}$ in (\ref{eq:truncD5D}) retains its form for $D\rightarrow5$, except for the first term 
$F_{(3)}$ which disappears.

The resulting Lagrangian (\ref{eq:L2DredS3}) after rescaling (\ref{eq:rescalingS3})
also yields a non-trivial smooth limit to $D\rightarrow5$, given by
\begin{equation}
{\cal L}_{2} =
 \rho\,\tilde{R}\,{\tilde\hodge 1}
+\tfrac14\,\rho\,{\tilde\hodge {\cal D}} \tilde{T}_{ij}^{-1}\wedge {\cal D} \tilde{T}_{ij} 
-\tfrac14\,\rho^{2}\,\tilde{T}_{ij}^{-1}\tilde{T}_{k\ell}^{-1} \,{\tilde\hodge F_{(2)}^{ik}}\wedge F_{(2)}^{j\ell} 
- \tfrac12\,g^2\left(2\,\tilde{T}_{ij}\tilde{T}_{ij}-\tilde{T}_{ii}^2 \right) {\tilde\hodge 1}
\;,
\label{eq:L2D2S3}
\end{equation}
again differing from the naive $D\rightarrow5$ limit of (\ref{eq:L2DredS3}) by the non-trivial powers of $\rho$.
Again, we may now directly prove that the regular reduction ansatz 
(\ref{eq:truncD5}) defines a consistent truncation of the 5-dimensional bosonic string (\ref{eq:S3L}) on $S^3$,
and we give the details below.
The resulting two-dimensional theory is given by the Lagrangian (\ref{eq:L2D2S3}).
For later convenience, we also note that the dual field strength (\ref{eq:F3Hodge}) for $D=5$ 
and in terms of the fields (\ref{eq:rescalingS3}) is found to be
\begin{align}
\hat{{\cal F}}_{(2)}\equiv&\;\;
 e^{-\sqrt{8/3}\, \hat\phi}\, {\hat \hodge  \hat F_{(3)}} 
 \label{eq:F3HodgeD5}\\
 =&\; 
 -g\, \tilde{U}\, \tilde\omega_{2}
 + g^{-1}\, \rho\,\tilde{T}^{-1}_{ij}\, 
{\tilde\hodge {\cal D} \tilde{T}_{jk}}\wedge (\mu^k\, {\cal D}\mu^i) 
-\tfrac12\,g^{-2}  \rho^2\, \tilde{T}^{-1}_{ik}\, \tilde{T}^{-1}_{j\ell}\, 
{\tilde\hodge  F_{(2)}^{ij}}\wedge {\cal D}\mu^k\wedge
{\cal D}\mu^\ell.
\nonumber
\end{align}
In terms of this dual field strength, the original $D=5$ theory (\ref{eq:S3L}) can be equivalently rewritten as
\bea
{\cal L}_5 &=& \hat R\; {\hat \hodge 1} - 
\tfrac12 \,{\hat \hodge d\hat\phi}\wedge d\hat\phi  
-\tfrac12 \,e^{\sqrt{8/3}\, \hat\phi}\, {\hat \hodge \hat {{\cal F}}_{(2)}}\wedge \hat {{\cal F}}_{(2)}
\;.
\label{eq:L5dual}
\eea

\section{Consistent $S^{D-2}$ reduction}
\label{sec:consistentS}

It is now straightforward to generalize the results of the last section in order to define 
the general consistent $S^{D-2}$ reduction of~(\ref{eq:LDintro}),
thereby establishing the fourth family of consistent truncations in (\ref{eq:Dp23}). Our starting point is the Einstein-Maxwell dilaton system obtained by dualizing\footnote{This is in line with the dual Lagrangians (\ref{eq:L4dual}), (\ref{eq:L5dual})
of which our construction is the natural generalisation.}
 the $d$-form field strength $\hat F_{(d)}$ into an abelian 2-form field strength $ \mathcal{\hat F}_{(2)}$,
 \bea
{\cal L}_D &=& \hat R\; {\hat \hodge 1} - 
\tfrac12 {\hat \hodge d\hat\phi}\wedge d\hat\phi  
-\tfrac12 \,e^{\frac{\sqrt{2\,(D-1)}}{\sqrt{D-2}}\hat\phi}\; {\hat \hodge \hat {\cal F}_{(2)}}\wedge \hat {\cal F}_{(2)}
\;.
\label{eq:LDD}
\eea
This is the Lagrangian obtained from the circle reduction of pure gravity in $(D+1)$ dimensions.

\subsection{Reduction ansatz and two-dimensional theory}
\label{sec:reduction}

Extrapolating from the previous findings, we can construct a consistent truncation of the $D$-dimensional theory (\ref{eq:LDD}) on the sphere $S^d=S^{D-2}$ 
by the following reduction ansatz
\bea
d\hat s_D^2 &=& \rho^{-\frac{2(d-1)}{d(d+1)}}\,\tilde{\Delta}^{\frac{d-1}{d}}\,d\tilde{s}_{2}^2 
+ g^{-2}\, \rho^{2/d} \tilde\Delta^{-1/d}\,  \tilde{T}^{-1}_{ij}\, 
{\cal D}\mu^i\, {\cal D}\mu^j 
\;,\label{eq:truncationD}\\
e^{\sqrt{\frac{2\,(d+1)}{d}}\, \hat\phi} &=& \rho^{-\frac{2(d-1)}{d}} \, \tilde\Delta^{-\frac{d+1}{d}}\;,
\nonumber\\
\hat{\cal F}_{(2)} &=& 
 -g\,\rho^{\frac{d-3}{d+1}}\, \tilde{U}\, \tilde\omega_{2}
 +\frac{1}{g}\, \rho\,\tilde{T}^{-1}_{ij}\, 
{\tilde\hodge {\cal D} \tilde{T}_{jk}}\wedge (\mu^k\, {\cal D}\mu^i) 
-\frac1{2g^2}\,\rho^{\frac{d+5}{d+1}}\,\tilde{T}^{-1}_{ik}\, \tilde{T}^{-1}_{j\ell}\, 
{\tilde\hodge  F_{(2)}^{ij}} \,{\cal D}\mu^k\wedge
{\cal D}\mu^\ell
\;.
\nonumber
\eea
The matrix $\tilde{T}_{ij}$ now is a symmetric positive definite $(d+1)\times(d+1)$ matrix of unit determinant, and the quantities
\bea
\tilde\Delta=\mu^i\tilde{T}_{ij}\mu^j\;,\quad
\tilde U\equiv2\,\tilde{T}_{ik}\tilde{T}_{jk}\,\mu^i\mu^j-\tilde\Delta\tilde{T}_{ii}\;,
\eea
are defined as above.
The inverse sphere radius $g$ appears as a coupling constant for the ${\rm SO}(d+1)$ covariant derivatives
just as in (\ref{eq:covD}) above. The corresponding two-dimensional ${\rm SO}(d+1)$ field strength $F_{\mu\nu}{}^{ij}$ is given by (\ref{eq:F2YM}).

It is straightforward, although lengthy, to substitute the ansatz (\ref{eq:truncationD}) 
into the $D$-dimensional equations of motion obtained from (\ref{eq:LDD}), and to verify that these consistently truncate
to the field equations of a two-dimensional theory. We find that these equations can be derived from the following two-dimensional Lagrangian
\bea
{\cal L}_{2} &=& 
\rho\,R\;\tilde\hodge 1
+\frac14\,\rho\;\tilde\hodge{\cal D} \tilde{T}_{ij}^{-1}\wedge{\cal D} \tilde{T}_{ij} 
-\frac14\,\rho^{\frac{d+5}{d+1}}\,\tilde{T}_{ij}^{-1}\tilde{T}_{k\ell}^{-1}\ \tilde\hodge F_{(2)}^{ik}\wedge F_{(2)}^{j\ell} 
\nonumber\\
&&{}
- \frac12\,g^2\,\rho^{\frac{d-3}{d+1}}\left(2\,\tilde{T}_{ij}\tilde{T}_{ij}-\tilde{T}_{ii}^2 \right) \tilde\hodge 1
\;,
\label{eq:L2result}
\eea
which in particular matches the above results 
(\ref{eq:L2D2S2}) and (\ref{eq:L2D2S3}) for $S^2$ and $S^3$, respectively,
as well as the structures found for the $S^8$ case in \cite{Ortiz:2012ib,Anabalon:2013zka,Bossard:2023jid}.
This is a two-dimensional dilaton gravity coupled to a gauged ${\rm SL}(d+1)/{\rm SO}(d+1)$ coset
space sigma model with a potential.
The two-dimensional Lagrangian (\ref{eq:L2result}) can be equivalently rewritten as
\bea
{\cal L}'_{2} &=& 
 \rho\,R\;\tilde\hodge 1
+\frac14\,\rho\;\tilde\hodge{\cal D} \tilde{T}_{ij}^{-1}\wedge {\cal D} \tilde{T}_{ij} 
+\frac14\,g\,{\cal Y}_{ij}\, F_{(2)}^{ij}
\nonumber\\
&&{}
- \frac12\,g^2\,\rho^{\frac{d-3}{d+1}}\left(2\,\tilde{T}_{ij}\tilde{T}_{ij}-\tilde{T}_{ii}^2 \right) \tilde\hodge 1
-\frac14\,g^2\,\rho^{-\frac{d+5}{d+1}}\, {\cal Y}_{ij} {\cal Y}_{k\ell}  \,\tilde{T}_{ik}\tilde{T}_{j\ell}
\;\tilde\hodge 1
\;,
\label{eq:L2result_dual}
\eea
upon introducing auxiliary scalar fields ${\cal Y}_{ij}={\cal Y}_{[ij]}$. The latter are related to the $\tilde{T}_{ij}$
by the (non-abelian) duality equations
\bea
{\cal D} {\cal Y}_{ij} 
&=&-
2 \,\rho\,\tilde{T}^{-1}_{k[i} \;\tilde\hodge {\cal D} \tilde{T}^{\vphantom{-1}}_{j]k}\,  
\;,
\label{eq:dualityYT}
\eea
and can be eliminated 
by virtue of their field equations
\bea
F_{(2)}^{ij} =g\,
  \rho^{-\frac{d+5}{d+1}}\,\tilde{T}_{ik}\tilde{T}_{j\ell} {\cal Y}_{k\ell}\;\tilde\hodge1
  \;.
  \label{eq:FY}
\eea

In order to prove consistency of the ansatz (\ref{eq:truncationD}), it is useful to first 
work out the Hodge dual of the $D$-dimensional 2-form field strength $\hat{\cal F}_{(2)}$ from (\ref{eq:truncationD}) as\footnote{
The $D$-dimensional volume form is given by $\hat\omega_D = \frac1{d!}\,
g^{-d} \rho^{\frac{2-d+d^2}{d(d+1)}}\,\tilde{\Delta}^{\frac{d-1}{d}}
\,
\tilde\omega_2 \, \wedge \,{\cal D} \mu^{i_1} \wedge \dots \wedge  {\cal D} \mu^{i_{d}}\,\mu^{i_{d+1}}\,
\epsilon_{i_1i_2\dots  i_{d+1}}$, with the totally
antisymmetric tensor $\epsilon_{i_1i_2\dots  i_{d+1}}$, and $\tilde\hodge\tilde\omega_2=1$.
}
\bea
e^{\sqrt{\frac{2\,(D-1)}{D-2}}\, \hat\phi}\;\hat\hodge  \hat{\cal F}_{(2)} &\!=\!& 
\epsilon_{i_1 i_2 \dots i_{d+1}}\, 
\Big( 
 \ft{1}{d!}\,g^{1-d}\,  \tilde{U}\, \tilde{\Delta}^{-2}\, 
  {\cal D}\mu^{i_1}\wedge \dots \wedge  {\cal D}\mu^{i_{d}}\, 
\mu^{i_{d+1}}
\label{eq:*F2}\\
&&\qquad\qquad \quad - \ft{1}{(d-1)!}\,g^{1-d}\, \tilde{\Delta}^{-2} \, \tilde{T}_{ji_{d+1}}\, {\cal D} \tilde{T}_{ki_{1}}\wedge
{\cal D}\mu^{i_2} \wedge\dots \wedge  {\cal D}\mu^{i_{d}} \,
 \mu^j\, \mu^k  
 \nonumber\\
&&\qquad\qquad \quad
-\ft{1}{2\,(d-2)!}\,g^{2-d}\, \tilde{\Delta}^{-1}\, \tilde{T}_{ji_{d+1}}\,\,F_{(2)}^{i_1 i_2} \wedge
{\cal D}\mu^{i_3} \wedge \dots \wedge  {\cal D}\mu^{i_{d}} \, \mu^j 
\Big)
\,,
\nonumber
\eea
for $d\geq2$.\footnote{For $d=1$, the formula degenerates to $\,e^{2 \hat\phi}\,\hat\hodge  \hat{\cal F}_{(2)} = 
\epsilon_{i_1 i_2}\, \tilde{\Delta}^{-2}\, 
\Big(  \tilde{U}\, 
  {\cal D}\mu^{i_1}\mu^{i_{2}} -   \, \tilde{T}_{ji_{2}}\, {\cal D} \tilde{T}_{ki_{1}}\mu^j\mu^k\Big)$.} After some lengthy algebra, one can then prove the 
$D$-dimensional field equations
\bea
d\Big(e^{\frac{\sqrt{2\,(D-1)}}{\sqrt{D-2}}\hat\phi}\;\hat\hodge  \hat{\cal F}_{(2)} \Big)=0\,,
\eea
as a consequence of (\ref{eq:*F2}). For the computation one only needs to focus on the terms involving one or two derivatives along the external directions. These contributions can be shown to cancel out entirely by repeatedly making use of the Schouten identity
\bea
\epsilon_{[i_1 i_2 \dots i_{d+1}}\, V_{j]}=0\,.
\label{eq:Schouten}
\eea
Next, by a similar computation one proves the Bianchi identity
\bea
d   \hat{\cal F}_{(2)} =0\,,
\label{eq:Bianchi}
\eea
for $\hat{\cal F}_{(2)}$ from (\ref{eq:truncationD}), which requires the two-dimensional fields to obey the field equations
obtained from (\ref{eq:L2result}).
Finally,\footnote{
Following the tradition set in \cite{Cvetic:2000dm}, we shall not explicitly 
consider the reduction of the $D$-dimensional Einstein equations in this paper. Their consistency has been confirmed in all the explicit solutions that have been examined. Furthermore, 
our ansatz (\ref{eq:truncationD}) agrees with all previously-established special cases for which the Einstein equations have been explicitly proven~\cite{Anabalon:2013zka,Bossard:2022wvi,Bossard:2023jid}.
} 
we can check the dilaton equations of the $D$-dimensional theory (\ref{eq:LDD})
\bea
{-(-1)^D}\,
\tfrac{\sqrt{2(D-2)}}{\sqrt{D-1}}\,d\Big( {\hat \hodge d\hat\phi} \Big)
&=&
 e^{\frac{\sqrt{2\,(D-1)}}{\sqrt{D-2}}\hat\phi}\; {\hat \hodge\hat{\mathcal F}_{(2)}}\wedge \hat{\mathcal F}_{(2)}\,,
\label{dilatonD}
\eea
and show that after some lengthy computation and heavy use of (\ref{eq:Schouten}), again they reduce to particular combinations of the two-dimensional field equations.
A useful identity in this computation is the expression for the Hodge dual
\bea
{\hat \hodge\big(\tilde{T}_{ij}\, \mu^i\, {\cal D}\mu^j\big)} &=& - \frac{1}{(d-1)!} \,g^{2-d}\, \rho^{\frac{d-3}{d+1}}\,\epsilon_{i_1 \dots i_{d+1}}\, \times
\nonumber\\
&&{}
\qquad \times\,
\tilde T_{i\ell}\, \mu^i\, \big( \tilde\Delta\, \tilde T_{i_1 \ell} -
\tilde T_{i_1 j}\, \tilde T_{k \ell}\, \mu^j\, \mu^k\big) \, \mu^{i_2}\,
{\cal D}\mu^{i_3}\wedge \dots \wedge {\cal D}\mu^{i_{d+1}}\wedge  \tilde\omega_{2} \,,
\label{idF1}
\eea
c.f.\ identity (25) in  \cite{Cvetic:2000dm}.
We have thus established that the reduction ansatz (\ref{eq:truncationD}) indeed represents a consistent truncation
of the theory (\ref{eq:LDD}) to the two-dimensional theory (\ref{eq:L2result}).

\subsection{Uplift to $D+1$}

As discussed above, the $D$-dimensional theory (\ref{eq:LDD})  which we have consistently truncated on $S^{D-2}$,
is itself the $S^1$ reduction of pure gravity in $(D+1)$ dimensions. It is thus a natural question to study the further uplift
of the reduction ansatz (\ref{eq:truncationD}) to $(D+1)$ dimensions. This can be achieved by the standard Kaluza-Klein
formula
\bea
d\breve s_{D+1}^2 = e^{-\sqrt{\tfrac{2}{d(d+1)}}\hat\phi}d\hat s_{D}^2+e^{\sqrt{\tfrac{2d}{d+1}}\hat\phi}\big(dz+\hat A_{(1)}\big)^2\,,
\label{eq:KKD+1}
\eea
where $z$ denotes the additional circle coordinate, and where we used $d=D-2$.
The uplift thus requires an explicit expression for the gauge potential defining the 2-form field strength as
\bea
d\hat A_{(1)} &=& \hat{\cal F}_{(2)}
\;.
\label{eq:dAF}
\eea
Indeed, we can integrate up $\hat{\cal F}_{(2)}$ from (\ref{eq:truncationD}) to the following expression
\bea
\hat A_{(1)} &=&
 \frac{1}{2g}\,{\cal Y}_{ij}\,\mu^i\,{\cal D}\mu^j
-\frac{1}{2g}\, \rho\,\mu^i \mu^j \,\tilde{T}^{-1}_{ki}\,
{\tilde\hodge {\cal D} \tilde{T}_{jk}} +\frac{2}{g(d+1)}\,\tilde{\hodge } d \rho
\;,
\eea
whose exterior derivative can be shown to satisfy (\ref{eq:dAF}) after
using the two-dimensional field equations. This yields another check on the Bianchi identity (\ref{eq:Bianchi}).
The uplift of (\ref{eq:truncationD}) to $(D+1)$ dimensions is then given by combining this ansatz with (\ref{eq:KKD+1}) into
\bea
d\breve s_{D+1}^2 = \tilde\Delta \,d \tilde{s}^2_{2}+g^{-2}\rho^{\tfrac{4d}{d(d+1)}}\,\tilde T^{-1}_{ij}\,\mathcal D\mu^i \mathcal D\mu^j+\rho^{-\tfrac{2(d-1)}{d+1}}\tilde\Delta^{-1}\,(dz+\hat A_{(1)})^2\,.
\label{eq:truncationD+1}
\eea
It represents a consistent truncation of $(D+1)$-dimensional Einstein gravity to the two-dimensional theory (\ref{eq:L2result}).\footnote{Note that for $d=1$, \eqref{eq:truncationD+1} defines a non-trivial consistent truncation of $D=4$ Einstein's gravity to two dimensions.}

In~\cite{Cvetic:2003jy}, the $(D+1)$-dimensional uplift of the system (\ref{eq:LDD}) was used to provide an elegant explanation for the consistent
truncation on $S^2$ found in \cite{Cvetic:2000dm}. From the  $(D+1)$-dimensional point of view, the $S^2$ reduction of (\ref{eq:LDD}) corresponds to
a standard Scherk-Schwarz reduction~\cite{Scherk:1979zr} on ${\rm SU}(2)$\,.
In contrast, here the reduction ansatz (\ref{eq:truncationD+1}) does not provide any immediate insight as to why this reduction is consistent,
rather it appears to represent a non-trivial truncation of pure gravity by itself.
We will come back to this in the conclusions.

\subsection{Cosmological term}
\label{subsec:cc}

Let us briefly study if the consistent truncation (\ref{eq:truncationD}) is compatible with the presence of a `cosmological term'
\begin{equation}
{\cal L}_{D,m} = -\frac12\,m^2\,e^{b\hat\phi}\,{\hat \hodge 1}\;,
\label{eq:cc}
\end{equation}
in $D$ dimensions.
We find that the previous reduction ansatz still gives rise to a consistent truncation, with all internal coordinates factoring out from the
$D$-dimensional field equations, if the dilaton power in (\ref{eq:cc}) is given by
\bea
b^2 &=&\frac{2\,(D-3)^2}{(D-1)(D-2)}\;.
\label{eq:b-cc}
\eea
For $D=4$ and $D=5$ this indeed coincides with the values extrapolated from the results of  \cite{Cvetic:2000dm}.
The effect of the cosmological term to the two-dimensional theory (\ref{eq:L2result}) is an additional term in the scalar potential given by
\bea
{\cal L}_m  &=&
 -  \frac12\,m^2\,\rho^{-\frac{d-3}{d+1}}\;\tilde\hodge 1
\;,
\label{eq:cc2}
\eea
with $d=D-2$.
It is interesting to note that the $D$-dimensional cosmological term (\ref{eq:cc}) with dilaton power (\ref{eq:b-cc}) is
not compatible with the further uplift of the theory to $(D+1)$ dimensions, i.e.\ it does not arise by the $S^1$ reduction
of a cosmological constant in $(D+1)$ dimensions.
This is an indication of the fact that the non-abelian ${\rm SO}(d+1)$ gauge group of the two-dimensional theory is not embedded
into the geometric ${\rm SL}(d+1)$ symmetry, arising in the toroidal reduction of $(D+1)$-dimensional gravity.
We will come back to this in the conclusions.

\section{Solutions}
\label{sec:solutions}

In this section, we construct some solutions of the two-dimensional theory (\ref{eq:L2result}) and 
work out their uplift to $D$ dimensions. In particular, we explore two-dimensional AdS$_2$ solutions, 
uplifting to higher-dimensional AdS$_2\times \Sigma$ geometries.

\subsection{Field equations}

In order to construct solutions of the two-dimensional theory, let us first spell out the field equations
derived from (\ref{eq:L2result_dual}). In addition to the first-order equations (\ref{eq:FY}) and (\ref{eq:dualityYT})
for the field strength and the auxiliary scalars, respectively, the scalar fields $\tilde{T}_{ij}$ satisfy
the second order field equations
\bea
{\cal D}_\mu\big( \rho\,\tilde{T}_{k(\!( i}^{-1}\,{\cal D}^\mu \tilde{T}^{\vphantom{-1}}_{j)\!) k}\big) &=& 
g^2\rho^{-\frac{d+5}{d+1}}\, \tilde{T}_{m\ell} \tilde{T}_{k(\!(i} {\cal Y}_{j)\!)m} {\cal Y}_{k\ell}  \,
+2\,g^2\rho^{\frac{d-3}{d+1}}\left(2\,\tilde{T}_{k(\!(i}\tilde{T}_{j)\!)k}-\tilde{T}_{(\!(ij)\!)} \tilde{T}_{kk} \right) 
\;.
\label{eq:EOMT}
\eea
Here, double brackets $(\!( \dots)\!)$ refer to traceless symmetrization of indices.
Furthermore, the Einstein and dilaton equations for (\ref{eq:L2result_dual}) take the form
\bea
0 &=& R+\frac14\,{\cal D}_\mu \tilde{T}_{ij}^{-1}\,{\cal D}^\mu \tilde{T}_{ij} 
- \frac{g^2}2\,\frac{d-3}{d+1}\,\rho^{-\frac{4}{d+1}}\left(2\,\tilde{T}_{ij}\tilde{T}_{ij}-\tilde{T}_{ii}^2 \right) +\frac{g^2}4\,\frac{d+5}{d+1}\,\rho^{-\frac{2(d+3)}{d+1}}\, {\cal Y}_{ij} {\cal Y}_{k\ell}  \,\tilde{T}_{ik}\tilde{T}_{j\ell}
\;,\nonumber\\
0 &=& \nabla_\mu\partial^\mu\,\rho 
+ \frac12\,g^2\,\,\rho^{\frac{d-3}{d+1}}\left(2\,\tilde{T}_{ij}\tilde{T}_{ij}-\tilde{T}_{ii}^2 \right) 
+\frac14\,g^2\,\,\rho^{-\frac{d+5}{d+1}}\, {\cal Y}_{ij} {\cal Y}_{k\ell}  \,\tilde{T}_{ik}\tilde{T}_{j\ell}
\;,\nonumber\\
0 &=& 
\nabla_{\mu} \partial_\nu \rho 
-\frac14\,\rho\,{\cal D}_\mu \tilde{T}_{ij}^{-1}\,{\cal D}_\nu \tilde{T}_{ij} 
-\frac12\,g_{\mu\nu}\, \Box\,\rho
+\frac18\,\rho\,{\cal D}_\mu \tilde{T}_{ij}^{-1}\,{\cal D}^\mu \tilde{T}_{ij} 
\;.
\label{eq:EinsteinDilaton}
\eea
In the following, we are going to construct particular solution to these equations.

\subsection{${\rm SO}(d+1)$ invariant solution}
\label{sec:sod+1}

Let us first consider solutions that preserve the entire ${\rm SO}(d+1)$ symmetry of the theory.
${\rm SO}(d+1)$ invariance requires
\bea
\tilde{T}_{ij}=\delta_{ij}\;,\quad Y_{ij}=0=A_\mu{}^{ij}
\;,
\label{eq:sod+1}
\eea
such that the only non-trivial fields in the theory are the dilaton $\rho$ and the two-dimensional metric.
With the domain-wall ansatz
\bea
d\tilde{s}^2 &=& -e^{2A(r)}\,dt^2 +dr^2
\;,
\eea
the Einstein field equations imply that
\bea
A(r)=A_0 + {\rm log}(\rho'(r))
\;.
\eea
Upon inserting this into (\ref{eq:EinsteinDilaton}), all remaining equations reduce to 
a single differential equation for the dilaton $\rho(r)$
\bea
\rho''(r) &=& \frac14\,(d^2-1)\,g^2\,\rho(r)^{\frac{d-3}{d+1}}
\;.
\label{eq:ode_rho}
\eea
In particular, this shows that ${\rm SO}(d+1)$ invariance is not compatible with a constant dilaton.
The general solution of equation (\ref{eq:ode_rho}) carries one integration constant $\alpha$
(apart from the trivial shift freedom $r\rightarrow r+c$), and can be given implicitly in terms
of hypergeometric functions as
\bea
\frac{r}{\alpha\rho} &=& _2F_1\!\left(\tfrac12,\tfrac12+\tfrac1{d-1},\tfrac32+\tfrac1{d-1};
-\tfrac14\,\alpha^2\,g^2\,(d+1)^2\,\rho^{2(d-1)/(d+1)}
\right)
\;.
\eea
In the limit $\alpha\rightarrow\infty$, we find the particular solution
\bea
\rho(r)=(gr)^{\frac{d+1}{2}}\;,\quad A(r)=A_0+\frac12\,(d-1)\,{\rm log}(r)
\;.
\eea
Using the reduction formulae (\ref{eq:truncationD}) (and setting $A_0=0$), we find the $D$-dimensional uplift 
of this solution as
\bea
d\hat s_{D}^2 &=& (gr)^{-\frac{d-1}{d}}\,\left( -r^{d-1}\,dt^2 +dr^2
+  r^{2} \,ds_{S^d}^2\right)
\;,\nonumber\\
e^{\frac{\sqrt{2\,(D-1)}}{\sqrt{D-2}}\hat\phi} &=& (gr)^{-\frac{d^2-1}{d}} \, \;,
\nonumber\\
F_{(2)} &=& 
 g^{\frac{d-1}{2}}\,  (d-1)\,r^{d-2}\, dt\wedge dr
\;.
\eea
For $d=8$, this reproduces the half-supersymmetric domain wall solution of~\cite{Boonstra:1998mp,Bergshoeff:2004nq},
corresponding to the ten-dimensional D0-brane near-horizon geometry.

\subsection{AdS$_2$ solutions with constant scalars and dilaton}

Here, we will search for solutions in which all scalars and the dilaton are constant, such that the two-dimensional metric becomes AdS$_2$.
In order to keep things simple, we restrict to the Cartan truncation, i.e.\ the further consistent truncation of (\ref{eq:L2result_dual}) to singlets under the Cartan ${\rm U}(1)^{[\frac{d+1}{2}]}$
subgroup of the ${\rm SO}(d+1)$ gauge group. This truncation keeps only $[\frac{d+1}{2}]$ vector fields and $[\frac{d}{2}]$ scalar fields among the $\tilde{T}_{ij}$, together with the dilaton $\rho$ and the two-dimensional metric. By construction, all the retained fields are neutral under the remaining ${\rm U}(1)^{[\frac{d+1}{2}]}$ gauge group. It is technically useful to distinguish the cases of even and odd $d$,
although the form of the resulting solutions is very similar.

\subsubsection{$d=2k+1$}

Let us parametrize the ${\rm U}(1)^{k+1}$ singlets within the scalar matrix $\tilde{T}_{ij}$ as
\bea
\tilde{T}_{ij} &=& h_0^{-\frac{2}{d+1}}\,\delta_{ij}\,h_{[\frac{i+1}{2}]}\;,\quad
i, j=1, \dots, d+1
\;,\nonumber\\
&&{}
h_{k+1}\equiv1\;,\quad
h_0\equiv \prod_{a=1}^{k} h_a
\;,
\label{eq:pT}
\eea
in terms of $k$ scalars $h_a>0$, $a=1, \dots, k$, such that ${\rm det}\,\tilde{T}_{ij}=1$\,.
Accordingly, we parametrize the matrix ${\cal Y}_{ij}$ as
\bea
{\cal Y} &=&  \big(
\mathbb{Y} \otimes \varepsilon 
\big)
\;,\quad
\mathbb{Y}_{\alpha\beta} \equiv \delta_{\alpha\beta}\,y_\alpha\;,\quad
\varepsilon\equiv 
\left(\begin{smallmatrix}0&1\\-1&0\end{smallmatrix}\right)
\;,\quad\alpha=1, \dots, k+1
\;.
\eea
in terms of $k+1$ scalars $y_\alpha$, and similarly for the field strengths ${\cal F}_{\mu\nu}{}^{ij}$\,.
Plugging all this into the field equations (\ref{eq:EOMT}), (\ref{eq:EinsteinDilaton}),
shows that all fields can be determined as a function of the free parameters $h_a$ and $\rho$ as
\bea
F_{(2)}^{\alpha} &=&-2\,g\,
  \rho^{-\frac{4}{d+1}}\, \tilde\omega_2\,h_\alpha^{3/2} \sqrt{1-h_\alpha+\sum_{a=1}^k h_a }
\;,
\qquad
\alpha=1, \dots, k+1\;,
\nonumber\\[1ex]
R&=& -8\,\rho^{-\frac{4}{d+1}}\,h_0^{-\frac{4}{d+1}}\,\Big(\sum_{a=1}^k h_a+\sum_{a<b} h_ah_b\Big) ~<~ 0
\;.
\label{eq:FRodd}
\eea
In particular, the two-dimensional metric is AdS$_2$.

\subsubsection{$d=2k$}

Similar to (\ref{eq:pT}), in this case we parametrize the ${\rm U}(1)^{k}$ singlets within the scalar matrix $\tilde{T}_{ij}$ as
\bea
\tilde{T}_{ij} &=& h_0^{-\frac{2}{d+1}}\,\delta_{ij}\,h_{[\frac{i+1}{2}]}\;,\quad
i, j=1, \dots, d+1
\;,\nonumber\\
&&{}
h_{k+1}\equiv1\;,\quad
h_0\equiv \prod_{a=1}^{k} h_a
\;,
\eea
in terms of $k$ scalars $h_a>0$, such that ${\rm det}\,\tilde{T}_{ij}=1$\,.
Accordingly, we parametrize the matrix ${\cal Y}_{ij}$ as
\bea
{\cal Y} &=&  \begin{pmatrix}
\mathbb{Y} \otimes \varepsilon & \\[-2ex]
& 0
\end{pmatrix}
\;,\quad
\mathbb{Y}_{ab} \equiv \delta_{ab}\,y_a\;,\quad
\varepsilon\equiv 
\left(\begin{smallmatrix}0&1\\-1&0\end{smallmatrix}\right)
\;,
\eea
in terms of $k$ scalars $y_a$.
Plugging all this into the field equations  (\ref{eq:EOMT}), (\ref{eq:EinsteinDilaton}),
yields the condition
\bea
\sum_{a=1}^k h_a &=& \frac12\;.
\label{eq:con12}
\eea
All other fields can then be determined as a function of the remaining free parameters $h_a$ and $\rho$ as
\bea
F_{(2)}^{a} &=&-2\,
  \rho^{-\frac{4}{d+1}}\,\tilde\omega_2\,h_0^{-\frac{4}{d+1}}\,h_a^{3/2} \, \sqrt{1-h_a}
\;,\nonumber\\[1ex]
R&=&- \rho^{-\frac{4}{d+1}}\,h_0^{-\frac{4}{d+1}}\,\Big(1+8\sum_{a<b} h_ah_b\Big) ~<~0
\;.
\label{eq:FReven}
\eea
Again, the two-dimensional metric is AdS$_2$.
For $d=8$, this reproduces the solution (\ref{eq:FReven}) reproduces the solutions of \cite{Anabalon:2013zka}.

\subsubsection{$d=2$}

Using the reduction formulae (\ref{eq:truncationD}), we can construct the $D$-dimensional uplift 
of the above solutions. They describe $D$-dimensional AdS$_2\times \Sigma_d$ backgrounds where
$\Sigma_d$ is a deformed $S^d$-sphere preserving only ${\rm U}(1)^{[\frac{d+1}{2}]}\subset {\rm SO}(d+1)$
of the isometries of the round sphere. Rather than going through the general case, let us illustrate the uplift
for the $d=2$ case, i.e.\ uplift the solution (\ref{eq:FReven}) to $D=4$ dimensions.

For $d=2$, the condition (\ref{eq:con12}) implies that $h_1=\frac12$, and the only free parameter in (\ref{eq:FReven}) is the constant dilaton $\rho$,
which we may absorb into a rescaling of the fields. After setting $g=\ell^{-1}$ and some further rescaling of fields, the four-dimensional solution takes the form
\bea
d\hat{s}_4^2 &=&\Delta_0^{1/2}\,ds^2_{{\rm AdS}_2}
+\ell^2\,\Delta_0^{-1/2}\,
\big(
2\, {\cal D} \mu^a {\cal D} \mu^a   +d \mu^3 d \mu^3  
\big)
\;,\nonumber\\
\hat{\cal F}_{(2)} 
&=&
\frac1{\ell}\,{\rm sin}^2\theta\,\tilde\omega_2 
+
\frac{\sqrt{2}}{\ell}\,{\Delta}_0^{1/2}\,{\rm cos}\,\theta\;
\hat\star\tilde\omega_2 
\;,\qquad
 e^{\sqrt3\,\hat\phi} =
2\,\Delta_0^{-3/2}
\;,
\eea
with $\Delta_0=(1+{\rm cos}^2\theta)$, and 
we have parametrized $\mu=\{{\rm sin}\,\theta\;{\rm cos}\,\phi,{\rm sin}\,\theta\;{\rm sin}\,\phi, {\rm cos}\,\theta\}$.
The metric and volume form of AdS$_2$ are denoted by $ds^2_{{\rm AdS}_2}$ and $\tilde\omega_2$, respectively,
with AdS radius $\ell$. Moreover, we have introduced
\bea
{\cal D} \mu^a=d\mu^a +  A\,\epsilon_{ab} \mu^b
\;,
\quad
a=1,2\;,\qquad
dA=\ell^{-2}\,\tilde{\omega}_2
\;.
\eea
Using (\ref{eq:truncationD+1}), we may further uplift this solution to a solution of pure gravity in $D=5$ dimensions,
given by the metric
\bea
d\breve s_5^2 &=&
\Delta_0\,ds^2_{{\rm AdS}_2}
+\ell^2\,
\big(
2\, {\cal D} \mu^a {\cal D} \mu^a   +d \mu^3 d \mu^3  
\big)
+
\Delta_0^{-1}\,\big(dz+\sqrt{2}\ell\,\epsilon_{ab}\, \mu^a {\cal D} \mu^b\big)^2
\;.
\label{eq:D5Kerr0}
\eea
Using an explicit parametrization of the AdS$_2$ metric in coordinates $\{v, r\}$, 
as well as for the 1-form $A$, the metric (\ref{eq:D5Kerr0}) 
takes the form
\bea
\frac1{\ell^2}\,d\breve s_5^2 &=&
\frac{1+{\rm cos}^2\theta}{\ell^2}\,\Big(2dvdr-\frac{r^2}{\ell^2} dv^2\Big)
+(1+{\rm cos}^2\theta)\,d\theta
+\gamma_{ij}\,(dx^i+k^i dv)(dx^j+k^j dv)
\;,\;\;
\label{eq:Kerr}
\eea
where we have defined
\bea
x^i=\{\phi,z\}\;,\quad
k^i=-\frac1{\ell^2}\,\{r,0 \}\;,\quad
\gamma_{ij}\,dx^idx^j=\frac{4\,{\rm sin}^2\theta}{1+{\rm cos}^2\theta}\,\big(d\phi + \tfrac1{2\sqrt{2}\ell}\,dz\big)^2 +\frac2{\ell^2}\,dz^2
\;.
\label{eq:Kerr1}
\eea
The metric (\ref{eq:Kerr}), (\ref{eq:Kerr1}) has isometry group ${\rm SO}(2,1)\times {\rm U}(1)^2$, with the first factor realized on AdS$_2$
and the two abelian factors realized by shifts of the periodic coordinates $\phi$ and $z$.  
Up to redefinition of $\phi$ and $z$, this precisely reproduces the 
near-horizon of the boosted Kerr string, c.f.\ \cite{Frolov:1987rj,Kunduri:2007vf,Kunduri:2008rs,Hollands:2010bfi}.
The two-dimensional theory (\ref{eq:L2result_dual}) for $d=2$ thus captures a consistent truncation around
this near horizon geometry.

\subsection{(A)dS$_2 \times S^d$ solutions of higher dimensional theory with cosmological term}

We have seen in section~\ref{subsec:cc} that the consistent truncation supports the presence of a `cosmological term' (\ref{eq:cc}) 
in $D$ dimensions, which changes the scalar potential of the two-dimensional theory.
We may thus for this case reexamine the existence of ${\rm SO}(d+1)$ invariant solutions (\ref{eq:sod+1}).
Evaluating the new scalar potential in presence of (\ref{eq:cc2}), we find that the theory admits a solution
with constant dilaton $\rho=\rho_0$, provided the coefficient of the cosmological term is given by
\bea
m^2 &=&  g^2\,\rho_0^{\frac{2(d-3)}{d+1}}(d^2-1) 
\;.
\eea
In this case, the two-dimensional curvature scalar is determined from (\ref{eq:EinsteinDilaton}) as
\bea
R &=& -g^2\,(d-1)\,(d-3)\, \rho_0^{-\frac{4}{d+1}}
\;.
\eea
For $d\ge4$, i.e.\ $D\ge6$, we thus find a geometry AdS$_2 \times S^d$ with the round sphere $S^d$, if the $D$-dimensional
theory carries a cosmological term with positive $m^2$. In contrast, for $d=2$, 
the consistent truncation supports a dS$_2 \times S^2$ solution. 
For $d=3$ (i.e.\ a reduction from $D=5$ dimensions on $S^3$), the theory admits a solution of type ${\rm Mink}_2\times S^3$,
living within the consistent truncation.

\section{Conclusions}
\label{sec:conclusions}

In this paper, we have completed the classification of \cite{Cvetic:2000dm}, 
and worked out the consistent truncation of $D$-dimensional Kaluza-Klein gravity (\ref{eq:LDD})
on an $S^{D-2}$ sphere to two dimensions.
We have given the two-dimensional Lagrangian and explicitly constructed several families of solutions
as well as their uplift to $D$ dimensions. In particular, we have identified within the consistent truncations
several solutions with AdS$_2$ geometry that uplift to different AdS$_2\times \Sigma$ geometries,
and notably the near-horizon geometry of the boosted Kerr string.

This construction realizes the fourth and last of the families listed in (\ref{eq:Dp23}), identified in \cite{Cvetic:2000dm}
as potentially consistent sphere truncations.
As discussed in the introduction, the existence of these truncations requires the embedding of
the sphere isometry group into the global symmetry group of the toroidally compactified theory. Generically, the toroidal reduction of (\ref{eq:LDintro}) on $T^d$  admits an $\mathbb{R}\times {\rm GL}(d)$ global symmetry,\footnote{GL($d$) is the standard geometric symmetry of toroidal reductions. The $\mathbb{R}$ factor refers to the scaling symmetry of the $p$-forms and the shift of the dilaton, which is already present in $D$ dimensions.}
however this symmetry is enhanced at particular values of $a$, which is realized for the four families of (\ref{eq:Dp23}).
While this symmetry enhancement is only a necessary condition for consistency of the sphere truncations, their existence
can be proven by explicit computation as has been done in \cite{Cvetic:2000dm} and the present paper.
With hindsight though, as has become apparent in recent years, 
all the consistent sphere truncations corresponding to (\ref{eq:Dpex}) and (\ref{eq:Dp23})
owe their existence to some underlying symmetry structure in Riemannian, generalized, and exceptional geometry.

For the $S^2$ reductions, i.e.\ the first family of (\ref{eq:Dp23}), it was shown in \cite{Cvetic:2003jy} that the
consistent truncation of \cite{Cvetic:2000dm} becomes more transparent after further uplift of the $D$-dimensional theory 
(\ref{eq:LDintro}) to pure gravity in $(D+1)$ dimensions (for $a$ given as in (\ref{eq:S2L})). 
In terms of this higher-dimensional theory, the truncation amounts
to a standard Scherk-Schwarz reduction \cite{Scherk:1979zr} on the group manifold ${\rm SU}(2)$.
In particular, this explains the symmetry enhancement
\bea
\mathbb{R} \times{\rm GL}(2) &\hookrightarrow& {\rm GL}(3)
\;,
\eea
of the toroidal reduction of (\ref{eq:LDintro}). The ${\rm SO}(3)$ isometry group of the sphere reduction is then naturally embedded  
into the enlarged symmetry group.
For the $S^3$ reductions, i.e.\ the second family of (\ref{eq:Dp23}), 
the consistent truncation requires the symmetry enhancement 
 \cite{Maharana:1992my}
\bea
\mathbb{R} \times {\rm GL}(3)
&\hookrightarrow&
\mathbb{R} \times{\rm O}(3,3)
\eea
of the toroidal reduction of (\ref{eq:LDintro})  (for $a$ given as in (\ref{eq:S3L})), 
such that the ${\rm SO}(4)$ isometry group of the sphere 
can be embedded into the enlarged symmetry group. 
The existence of the consistent truncation can be attributed to the generalized parallelizability of this background 
within the double field theory formulation of (\ref{eq:LDintro})~\cite{Lee:2014mla,Hohm:2014qga,Baguet:2015iou}.
For the third family of (\ref{eq:Dp23}), i.e.\ the truncation on $S^{D-3}=S^d$,
the further symmetry enhancement  
after toroidal reduction to three dimensions \cite{Ehlers:1957,Sen:1994wr,Cremmer:1999du}
\bea
\mathbb{R} \times {\rm GL}(d)
&\hookrightarrow&
\mathbb{R} \times{\rm O}(d,d)
~\hookrightarrow~{\rm O}(d+1,d+1)
\eea
allows to embedd the ${\rm SO}(d+1)$ isometry group of the sphere 
into the enlarged symmetry group This consistent truncation then has a natural explanation
in the framework of the enhanced double field theory of \cite{Hohm:2017wtr}.

In the same spirit, the existence of the consistent truncation on $S^{D-2}$ constructed in this paper, allows 
for a natural explanation in the framework of the affine exceptional field theory 
of \cite{Bossard:2017aae,Bossard:2018utw,Bossard:2021jix},
as discussed in detail in \cite{Bossard:2022wvi,Bossard:2023wgg,Bossard:2023jid}.
Its starting point is the symmetry enhancement after toroidal reduction of (\ref{eq:LDintro}) 
to two dimensions (for $a$ given as in \eqref{eq:LDD}) according to \cite{Geroch:1972yt,Belinsky:1971nt,Maison:1978es,Julia:1982gx}
\bea
\mathbb{R} \times{\rm GL}(d) &\hookrightarrow& {\rm GL}(d+1)
~\hookrightarrow~   \mathbb{R} \ltimes\widehat{{\rm SL}(d+1)}
\;,
\label{eq:affine}
\eea
where $\widehat{{\rm SL}(d+1)}$ denotes the affine extension of ${\rm SL}(d+1)$.
We have observed in section \ref{subsec:cc} that the addition of a cosmological term is compatible with the consistent truncation
but obstructs the uplift of (\ref{eq:LDintro})  to $D+1$ dimensions. This is a manifestation of the fact that the ${\rm SO}(d+1)$ isometry
group of the sphere is not embedded into the intermediate ${\rm GL}(d+1)$. This was already noticed in \cite{Bossard:2022wvi,Bossard:2023jid}, to which we refer for more details.

An immediate application of consistent truncations is the construction of higher-dimensional solutions. By construction,
any solution of the lower-dimensional theory uplifts into a solution of the higher-dimensional theory. In particular, solutions with
constant scalar fields, that take a simple form in the lower-dimensional theory may give rise to higher-dimensional backgrounds with complicated internal geometry, which has been exploited in many instances in the past.
As an illustration, we have constructed a few solutions of the two-dimensional theory (\ref{eq:L2result}) 
together with their higher-dimensional uplift, but it would certainly be very interesting to generalize this to a more systematic and exhaustive construction of solutions. Within the Cartan truncation, it should be straightforward to work out the general rotating brane solutions as in \cite{Cvetic:1999xp,Harmark:1999xt,Anabalon:2013zka} and study their thermodynamic properties using Sen's entropy function formalism \cite{Sen:2005wa}.
Going beyond the Cartan truncation of the scalar sector will require to deal with non-trivial non-abelian gauge fields and may lead to entirely new classes of higher-dimensional solutions.

As another application of consistent truncations, these provide valuable tools in the context of holography, since they allow to perform supergravity calculations directly within the lower-dimensional theory. E.g.\ this offers an immediate path to the computation of conformal dimensions and correlation functions of the operators dual to the fields of the consistent truncation. We have shown that the two-dimensional Lagrangian (\ref{eq:L2result}) carries a solution that uplifts to the near-horizon geometry of the boosted Kerr string (\ref{eq:Kerr}). It will be very interesting to systematicaly analyze within this theory the perturbations around this background in the context of the Kerr/CFT correspondence~\cite{Guica:2008mu}, see e.g.~\cite{Murata:2011my} for a related study.
More recently, it has been shown that consistent truncations together with the underlying exceptional geometry may even allow to access the full Kaluza-Klein spectrum, i.e.\ the infinite towers of fluctuations, around a given background \cite{Malek:2019eaz,Malek:2020yue}.
It would be highly interesting to develop similar technology around the backgrounds considered here, and beyond, with potential applications to compactifications of maximal supergravity, see e.g.~\cite{Michelson:1999kn,Corley:1999uz,Lee:1999yu,Larsen:2014bqa}. We hope to come back to this in the future.

\subsection*{Acknowledgements}

We would like to thank Guillaume Bossard, Gianluca Inverso, Axel Kleinschmidt and Hermann Nicolai for useful discussions. This work has received funding from the European Research Council (ERC) under the European Union’s Horizon 2020 research and innovation programme (grant agreement No 740209).


\providecommand{\href}[2]{#2}\begingroup\raggedright\endgroup

\end{document}